\documentclass[prd,nofootinbib,eqsecnum,final]{revtex4-2}
  \usepackage{bm}
   \usepackage{amsmath,extarrows}
    \usepackage{amssymb}
     \usepackage{pifont}
\usepackage{graphicx,color}
\usepackage{slashed}

\newcommand{\nn}{\nonumber}

\newcommand{\tr}{\text{tr}}

\renewcommand{\(}{\left(}
\renewcommand{\)}{\right)}
\renewcommand{\[}{\left[}
\renewcommand{\]}{\right]}

\newcommand{\fnot}[1]{\slashed{#1}}

\begin{document}

\preprint{IPARCOS-UCM-26-057}

\title{Collins-Soper kernel at large $b$ from Chiral Perturbation Theory}

\author{Alexey Vladimirov}
\thanks{\href{mailto:alexeyvl@ucm.es}{alexeyvl@ucm.es}}
\affiliation{Departamento de Física Teórica and IPARCOS, Universidad Complutense de Madrid, \\ Plaza de Ciencias 1, 28040 Madrid, Spain}

\begin{abstract}

We construct effective operators for transverse-momentum-dependent (TMD) distributions in Chiral Perturbation Theory. The operators constructed from pion and nucleon fields possess the main properties of the QCD operators, including multiplicatively renormalizable rapidity divergences. Extracting the leading rapidity divergence, we find the expression for the Collins-Soper kernel at large $b$ and at the low-energy normalization point, which is model-independent up to an additive constant. In this regime, the Collins-Soper kernel has a nearly constant behavior. We also find that at large $b$ it depends on the isospin channel and on the hadron type, although this dependence is numerically very small.
\end{abstract}

\maketitle 

\section{Introduction}

The transverse momentum dependent (TMD) factorization theorem \cite{Collins:2011zzd, Echevarria:2011epo} is one of the most established approaches to study the internal structure of the hadrons and describe the transverse momentum differential cross-sections. Within this approach, the hadron structure is encoded in the TMD distributions -- parton distribution functions (PDFs) and fragmentation functions (FFs) -- that are generally interpreted as distributions of partons with respect to their momentum fraction $x$ and transverse momentum $p_T$ (or $b$ in the position space). Various regions of these functions possess different properties, in particular the small-$b$ regime can be directly computed in QCD in terms of collinear distributions (for example, \cite{Echevarria:2016scs, Ebert:2020yqt, Rein:2022odl}). At large $b$, on the contrary, perturbative QCD cannot give any explicit information about their structure and thus, in this regime, they remain subject to fitting and modeling. In this work, we demonstrate how one can construct analogs of TMD distributions in Chiral Perturbation Theory (ChPT) \cite{Weinberg:1978kz, Gasser:1983yg}, and deduce various properties from this construction.

The TMD distributions are universal objects well defined in QCD. They could be extracted from data (see for instance \cite{BermudezMartinez:2022ctj, Moos:2023yfa, Moos:2025sal, Bacchetta:2025ara, Bacchetta:2024qre}) or in lattice QCD simulations (see for instance \cite{Shanahan:2021tst, LatticePartonLPC:2022eev, Avkhadiev:2024mgd, Bollweg:2024zet, Tan:2025ofx}). However, all these studies have rather weak precision in the large-$b$ region (which loosely corresponds to the small transverse-momentum region). There are two reasons for this. The first one is that the data, as well as lattice simulations, are very sparse and noisy in this regime, and thus do not constrain the fit parameters sufficiently well. The second one is that QCD is essentially non-perturbative in this regime, and does not provide any hint or model-independent statement about the functional behavior of TMD distributions. There exist various model computations and estimates of the large-$b$ behavior, based, for instance, on dynamical chiral symmetry breaking \cite{Strikman:2009bd, Schweitzer:2012hh} or the QCD instanton vacuum \cite{Liu:2024sqj}, as well as on other approaches \cite{Bacchetta:2008af, Pasquini:2008ax, Collins:2014jpa, Noguera:2015iia, Scimemi:2016ffw}. However, they inevitably rely on model assumptions. Together, these factors lead to a rather disastrous situation, in which extractions of TMD distributions based on different analyses lead to very distinct curves at large $b$, although they agree at small $b$ (see comparison plots in refs.~\cite{Moos:2025sal, Chang:2026rft, Vladimirov:2020umg}). In these circumstances, any model-independent statement about large-$b$ behavior can significantly improve the quality of extractions.

A particularly interesting possibility to investigate the non-perturbative regime of QCD is to utilize the low-energy effective field theories, such as ChPT, which describes the strong interaction at scales $\mu\lesssim 0.4$\,GeV. In the position space this corresponds to distances $b\gtrsim 2.5$\,GeV$^{-1} \sim 0.5$\,fm, exactly the regime where the models suffer the most. However, the application of ChPT is not that straightforward since it requires the construction of an analog of the QCD operator with effective degrees of freedom. In other words, one should build an operator relation
\begin{eqnarray}\label{intro:matching}
O[\bar q,q,A_\mu] \longrightarrow \sum_n c^{(n)} O_n[\bar N,N,\pi],
\end{eqnarray}
where the TMD operator on the left is made of QCD fields, and the TMD operator on the right is made of ChPT fields. These operators should have the same properties, such as quantum numbers, discrete and topological symmetries. The summation runs over various possible operators in ChPT, with the same properties, each of which is accompanied by its own low-energy constant (LEC). There is an infinite number of possible effective operators, however, their dimension grows and in the low-energy limit only a few contribute. This allows one to investigate various QCD observables in the low-energy regime in a model-independent way.

Traditionally, the effective field theory approach has been used for local operators, but it can also be applied to non-local light-cone partonic observables. The central idea is that the operator matching (\ref{intro:matching}) can be done for non-local light-cone operators preserving their twist, as a fundamental property of any partonic observable. In ChPT it implies a modification of the counting rules \cite{Kivel:2002ia} as
\begin{eqnarray}\label{def:counting}
(n\cdot \partial)\sim \mathcal{O}(1),
\end{eqnarray}
instead of the usual $\partial \sim m_\pi$. Such an extension does not violate the counting rules of ChPT itself, since $n^2=0$ and the vector $n$ does not enter the action.  In this way one can construct the whole tower of local operators with given twist, and sum them into non-local form, with LECs $c^{(n)}$ promoted to generating functions. This was first demonstrated in ref.~\cite{Polyakov:1998ze} for the pion distribution amplitude, and afterwards generalized to the case of Generalized Parton Distributions (GPDs) \cite{Kivel:2002ia, Diehl:2005rn, Kivel:2008ry, Perevalova:2011qi, Moiseeva:2012zi} and other partonic observables \cite{Granados:2013moa, Salamu:2018cny, Salamu:2019dok}. In the case of GPDs, the application of ChPT allows the determination of the small-$\Delta$ (inter-state momentum transfer) dependence and of the explicit dependence on the pion mass, which is used in the lattice extrapolations. Recently, similar ideas were applied to the soft-collinear effective field theory (SCET) basis elements \cite{Feldmann:2026tex}.

There were also attempts to apply a similar approach to TMD operators \cite{Copeland:2024wwm, Copeland:2024cgq}, which found that certain properties of TMD operators are absent in ChPT. However, and this is one of the principal arguments of this work, the construction of the effective TMD operator must follow different criteria in comparison to collinear operators. Most importantly, it must obey the TMD evolution, which is a consequence of its topology and algebra. Namely, the leading-twist TMD operator is a composite operator made of two semi-compact operators of twist-1 \cite{Vladimirov:2021hdn, Rodini:2022wki} (the so-called TMD-twist-(1,1) operator). Here, we construct such operators in ChPT and demonstrate that they obey all main properties of those in QCD. In particular, they have rapidity divergences and allow the determination of the Collins-Soper (CS) kernel in ChPT.

The paper has the following structure. In sec.~\ref{sec:construct}, we review the main properties of TMD operators in QCD, and discuss how they can be realized in ChPT. In particular, in sec.~\ref{sec:WL}, we construct the chiral Wilson line, which realizes the parallel transport in the background pion field. The TMD operators are then constructed in sec.~\ref{sec:nucleon} and sec.~\ref{sec:pion}, for the nucleon and pion cases, respectively. The chiral loop corrections are analyzed in sec.~\ref{sec:loop}, where we show that TMD operators in ChPT have rapidity divergences, which appear at two loops, and compute them. In sec.~\ref{sec:sf}, we construct the renormalization factor for rapidity divergences in the usual form of a soft factor made of chiral Wilson lines, and demonstrate that it cancels the rapidity divergences. Finally, in sec.~\ref{sec:discussion}, we summarize the findings, derive the large-$b$ form of the CS kernel, which is the main result of the paper, and discuss its peculiarities.

\section{Constructing effective TMD operators}
\label{sec:construct}

The initial goal is to construct an analog of the TMD operator in the framework of ChPT. This procedure is well elaborated for local operators (see, for instance, \cite{Diehl:2005rn, Arndt:2001ye, Chen:2001pva}), and also for light-cone collinear operators (see, for instance, \cite{Kivel:2002ia, Kivel:2008ry, Moiseeva:2012zi}). TMD operators have their own specifics: they are products of two semi-compact operators, and this structure, rather than the global quantum numbers, defines their properties. In this section, we describe these elements and their consequences in QCD, and review the main elements of TMD physics. Then we review the available building blocks of ChPT and construct auxiliary elements, which are then used to compose TMD operators in the following sections.

\subsection{TMD operators: general discussion}

The TMD factorization theorem expresses the cross-section in terms of TMD distributions, which in turn are forward matrix elements of TMD operators
\begin{eqnarray}\label{def:op->TMD}
\Phi^{[\Gamma]}(x,b)=\int \frac{dz}{2\pi} e^{-ixp^+z}\langle p,s|\mathcal{O}^{[\Gamma]}(z,b)|p,s\rangle ,
\end{eqnarray}
where $x$ is the parton momentum fraction, $b$ is the transverse distance Fourier conjugated to transverse momentum, and $\Gamma$ refers to various quantum numbers of the operator. For brevity, we do not indicate the renormalization scales: the scale $\mu$ is the ultraviolet (UV) renormalization scale, and the scale $\zeta$ is the rapidity renormalization scale. With respect to both scales the TMD operator is multiplicatively renormalizable \cite{Vladimirov:2017ksc}, and thus gives rise to a pair of evolution equations \cite{Aybat:2011zv, Chiu:2012ir} of the form
\begin{eqnarray}\label{evol:mu}
\mu^2\frac{d}{d\mu^2}\Phi^{[\Gamma]}(x,b)&=&\frac{1}{2}\gamma_F(\mu,\zeta)\Phi^{[\Gamma]}(x,b),
\\\label{evol:zeta}
\zeta\frac{d}{d\zeta}\Phi^{[\Gamma]}(x,b)&=&-\mathcal{D}(b,\mu)\Phi^{[\Gamma]}(x,b),
\end{eqnarray}
where $\gamma_F$ is the UV anomalous dimension, and $\mathcal{D}$ is the CS kernel, which is sometimes called the rapidity anomalous dimension. Both these anomalous dimensions are determined by the corresponding (UV and rapidity) divergences. Generally speaking, both $\gamma_F$ and $\mathcal{D}$ carry labels corresponding to quantum numbers, and are distinct for various TMD distributions, although they form large universality families. For instance, the anomalous dimensions of leading-twist TMD distributions differ only by the color representation, i.e., there are universal quark and universal gluon anomalous dimensions.

The TMD operators are ordered by TMD-twist -- a pair of numbers $(N_1,N_2)$ \cite{Vladimirov:2021hdn, Rodini:2022wki}. Each unique pair corresponds to a set of operators autonomous under the TMD evolution, i.e. which do not mix within equations (\ref{evol:mu}, \ref{evol:zeta}). Each TMD-twist possesses its own pair of $\gamma_F$ and $\mathcal{D}$. This is a defining property of TMD operators, which should be preserved in any construction, because it keeps the definition of a distribution clear, in the sense that it prevents uncontrollable mixing with other operators. The notion of TMD-twist is based on the Lorentz invariance, similarly to the geometrical twist \cite{Gross:1971wn}, which defines irreducible families of collinear distributions. In this work, we are interested only in the TMD distributions that appear at leading power, which all belong to TMD-twist-(1,1), which is often referred to as twist-two TMD operator. The details of their construction in QCD are presented below.

The TMD-twist-(1,1) operator is a product of two semi-compact operators of geometrical twist-1 separated by a transverse distance $b$
\begin{eqnarray}
\mathcal{O}_{11}(z,b)\sim\overline{U}_1(z,b)\dots U_1(0,0),
\end{eqnarray}
where dots denote possible algebraic structures, such as Dirac, color or flavor matrices, which do not affect the renormalization properties of TMD operator, but define its quantum numbers. The operator $U(z,b)$ is a semi-compact operator of geometric twist-one (i.e. dimension minus spin equals 1). It corresponds to the leading twist part of the propagator of a parton from the light-cone infinity\footnote{We use the standard notation for the light-cone decomposition of a vector $v^\mu =v^+ \bar n^\mu+v^- n^\mu+v_T^\mu$ with $n^2=\bar n^2=0$ and $(n\bar n)=1$.} $\pm \infty n+b$ to the point $zn+b$. The direction of the infinity is defined by the process, i.e. it is $+\infty$ for semi-inclusive deep inelastic scattering, and $-\infty$ for the Drell-Yan reaction. In the text we omit the $\pm$ sign of $\infty$ for brevity, except for the few occasions when it is important.

In QCD such operator consists of a Wilson line attached to the ``good component'' (i.e. the component with the maximum spin projection) of the parton field. For example, for the quark it reads
\begin{eqnarray}\label{def:tw1-QCD}
U_{1f}(z,b)=W_T\times[\infty n+b,zn+b]\xi_f(zn+b),
\end{eqnarray}
where $f$ is the flavor index, $\xi= \gamma^-\gamma^+q/2$ is a good projection of the quark field, and the QCD gauge link is
\begin{eqnarray}
[x_1,x_2]=P\exp\(-ig\int_{x_1}^{x_2} dx^\mu A_\mu(x)\).
\end{eqnarray}
The factor $W_T$ is a transverse link \cite{Belitsky:2002sm}, which connects the point $\infty n+b$ to the transverse infinity and makes the operators formally gauge invariant. 

The operator (\ref{def:tw1-QCD}) can be seen as the propagator of the quark in the background of the strong fields within the hadron. Importantly, along its full path, from the point of creation of the hadron, $\infty n$, to the measurement point $zn$, it preserves its chirality.

The operator (\ref{def:tw1-QCD}) is formally of twist-1. To see this, one regularizes $\infty n \to Ln$, and performs the local expansion\footnote{Note that the twist of $W_T$ is not very well defined. Thus, the formal expansion (\ref{def:formal-expansion}) implies the use of a non-singular gauge, where one can safely set $W_T\sim 1$, and consider it just as a factor that absorbs the gauge transformation. In this way, each term of the series (\ref{def:formal-expansion}) is gauge-invariant.}
\begin{eqnarray}\label{def:formal-expansion}
U_{1f}(0,0)=\lim_{L\to \infty}\sum_{n=0}^\infty \frac{(-L)^n}{n!}W_T(Ln)\times(n\cdot D)^n\xi_f(Ln)
=\lim_{L\to \infty}\sum_{n=0}^\infty \frac{(-L)^n}{n!}U^{(n)}_{1f}(Ln),
\end{eqnarray}
where $D_\mu=\partial_\mu-igA_\mu$ is the QCD covariant derivative. Although this formulation looks formal, it captures the physics properly, allowing one, for instance, to compute the operator product expansion beyond leading power and twist, and to observe the sign-change relation \cite{Moos:2020wvd}.  

In this work, we consider only quark TMD distributions, which thus carry spinor and flavor indices. For convenience, these indices are projected to the Dirac structure in the following manner 
\begin{eqnarray}\label{semi-compact->TMD}
\mathcal{O}^{[\Gamma]}_{fg}(z,b)=\overline{U}_{1f}(z,b)\frac{\Gamma}{2}U_{1g}(0,0),
\end{eqnarray}
where $\Gamma\in \{\gamma^+,\gamma^+\gamma^5,i\sigma^{\alpha+}\gamma^5\}$. The standard parametrization of TMD distributions is then \cite{Mulders:1995dh}
\begin{eqnarray}\label{def:+}
\Phi^{[\gamma^+]}(x,b)&=&f_1(x,b)+i \epsilon^{\mu\nu}_Tb_\mu s_{T\nu} M f_{1T}^\perp(x,b),
\\\label{def:+5}
\Phi^{[\gamma^+\gamma^5]}(x,b)&=&s_L g_1(x,b)+i (b\cdot s_T) M g_{1T}^\perp(x,b),
\\\label{def:a+5}
\Phi^{[i\sigma^{\alpha+}\gamma^5]}(x,b)&=&s_T^\alpha h_1(x,b)-is_L b^\alpha Mh_{1L}^\perp(x,b)+i\epsilon_T^{\alpha \mu}b_\mu M h_1^\perp(x,b) \\\nn && \qquad
-\frac{b^2M^2}{2}
\(\frac{g_T^{\alpha \mu}}{2}-\frac{b^\alpha b^\mu}{b^2}\)s_{\mu T}h_{1T}^\perp(x,b),
\end{eqnarray}
where $\epsilon^{\mu\nu}_T=\epsilon^{\mu\nu\alpha\beta}\bar n_\alpha n_\beta$, $s^\mu$ is the spin-vector with $s_L$ and $s^\mu_T$ being its longitudinal and transverse projections. These TMD distributions are independent, but obey the same evolution equations due to the fact that the renormalization structure of the TMD operator is independent of $\Gamma$.

One of the predictions of QCD is that the CS kernel is independent of the flavor and the type of hadron, and depends only on the color representation of the Wilson line. This prediction, however, should be taken carefully, because it is valid only in perturbative QCD (although at all orders of perturbative QCD). Therefore, at low energy, below the confinement and chiral phase transition threshold this prediction may be violated. We discuss it in detail below.

\subsection{Elements of Chiral perturbation theory}

The next task is to construct the analog of the semi-compact operator (\ref{def:tw1-QCD}) in ChPT. We consider the pion-nucleon ChPT in the relativistic form. This theory describes \cite{Gasser:1987rb, Bernard:1995dp, Scherer:2002tk} the strong interaction at low energy via the nucleon field $N$ and the pion fields $\pi^a$. Its leading order Lagrangian (in the isospin limit) is
\begin{eqnarray}\label{def:L}
\mathcal{L}&=&\frac{F_\pi^2}{4}\mathrm{tr}\(\partial_\mu U\partial^\mu U^\dagger\)+\frac{F_\pi^2}{4}m_\pi^2\mathrm{tr}\(U^\dagger+U\)
+\overline{N}\(i\fnot{\partial}+i\gamma^\mu\Gamma_\mu-M+\frac{g_A}{2}\gamma^\mu\gamma^5u_\mu\)N,
\end{eqnarray}
where $F_\pi$ is the pion decay constant, $M$ is the nucleon mass, $m_\pi$ is the pion mass and $g_A$ is the axial coupling. The field $N$ is the nucleon field, and $U$ is the chiral field made of pions,
\begin{eqnarray}\label{def:U}
U=u^2=\exp\(\frac{i\pi^a\tau^a}{F_\pi}\),
\end{eqnarray}
where $\tau^a$ are the Pauli matrices. The pion fields enter the nucleon part of the Lagrangian via the chiral connection and the chiral vielbein defined as
\begin{eqnarray}\label{def:Gamma}
\Gamma_\mu&=&\frac{1}{2}\(u^\dagger\partial_\mu u+u\partial_\mu u^\dagger\)=\frac{i}{4F_\pi^2}\epsilon^{abc}\pi^a\partial_\mu \pi^b\tau^c+\mathcal{O}(\pi^4),
\\\label{def:umu}
u_\mu&=&i\(u^\dagger\partial_\mu u-u\partial_\mu u^\dagger\)=-\frac{\tau^a\partial_\mu\pi^a}{F_\pi}+\mathcal{O}(\pi^3).
\end{eqnarray}

The Lagrangian (\ref{def:L}) is invariant under the global chiral group $SU(2)_R\times SU(2)_L$. In QCD, its elements $(R,L)$ act on the chiral projections of the quark fields as 
\begin{eqnarray}\label{def:chiral-transform}
q_R\to R q_R,\qquad q_L\to Lq_L,\qquad \text{with}~~q_{R,L}=\frac{1\pm\gamma^5}{2}q.
\end{eqnarray}
In nucleon-pion ChPT, this symmetry is realized non-linearly \cite{Coleman:1969sm, Callan:1969sn}
\begin{eqnarray}\label{sym:chiral}
U(x)\to RU(x)L^\dagger,\qquad u(x)\to Ru(x)h^\dagger(x)=h(x)u(x)L^\dagger,\qquad N(x)\to h(x)N(x),
\end{eqnarray}
where $h(x)\in SU(2)$ is the compensator field. It is a function of $R$, $L$ and $\pi(x)$, and thus it is local, even though the chiral transformation (\ref{def:chiral-transform}) is global.

\subsection{Wilson lines in ChPT}
\label{sec:WL}

The field combinations (\ref{def:Gamma}, \ref{def:umu}) transform as
\begin{eqnarray}\label{sym:blocks}
\Gamma_\mu\to h\Gamma_\mu h^\dagger+h\partial_\mu h^\dagger,\qquad u_\mu\to hu_\mu h^\dagger.
\end{eqnarray}
Thus, $\Gamma_\mu$ transforms as a gauge field of the local group $h(x)$, while $u_\mu$ transforms covariantly. Using these elements, we construct two auxiliary fields
\begin{eqnarray}\label{def:A}
A_\mu=\Gamma_\mu +\frac{ia}{2}u_\mu=\frac{1-a}{2}u^\dagger\partial_\mu u+\frac{1+a}{2}u\partial_\mu u^\dagger, \qquad \widetilde{A}_\mu=\Gamma_\mu -\frac{ia}{2}u_\mu=\frac{1+a}{2}u^\dagger\partial_\mu u+\frac{1-a}{2}u\partial_\mu u^\dagger.
\end{eqnarray}
The parameter $a$ is real. It is an additional parameter of our construction, which is related to the hadronic environment and is fixed later on. Under the chiral transformation, the combination $A_\mu$ transforms as $\Gamma_\mu$, i.e.,
\begin{eqnarray}\label{A->hAh+hdh}
A_\mu\to h A_\mu h^\dagger+h\partial_\mu h^\dagger.
\end{eqnarray}
and the same holds for $\widetilde{A}_\mu$. Thus, both can be seen as gauge fields.

The reason to introduce two fields is that they are related by the parity transformation $A_\mu(x) \xleftrightarrow{P}\widetilde{A}^\mu(\tilde x)$, with $\tilde x=(x^0,-\vec x)$, which is used in what follows. With the help of these fields one immediately builds the covariant derivatives as
\begin{eqnarray}
D_\mu=\partial_\mu+A_\mu,\qquad \widetilde{D}_\mu=\partial_\mu+\widetilde{A}_\mu,
\end{eqnarray}
such that $D_\mu N\to h D_\mu N$ and $\widetilde{D}_\mu N\to h \widetilde{D}_\mu N$. With the help of these covariant derivatives we can build the chiral analogs of the operators $U^{(n)}_{1f}(0)$ (\ref{def:formal-expansion}). Namely,
\begin{eqnarray}\label{op:URn}
U^{(n)}_{1R}(x)&\xrightarrow{\text{ChPT}}& u(x) (n\cdot D)^n N(x),
\\\label{op:URl}
U^{(n)}_{1L}(x)&\xrightarrow{\text{ChPT}}& u^\dagger(x) (n\cdot \widetilde{D})^n N(x),
\end{eqnarray}
where the indices $R$ and $L$ indicate the right and left projections of the quark fields (\ref{def:chiral-transform}) and the flavor indices are implied. Note that each effective operator should be accompanied by a unique LEC. All these operators have the same chiral counting order, thanks to the rule (\ref{def:counting}). 

The summation of the tower of operators (\ref{op:URn}, \ref{op:URl}) can be written as (here for the right operator)
\begin{eqnarray}\label{op:summing-U}
\sum_{n=0}^\infty \frac{(-L)^n}{n!}c_{U_R}^{(n)} u(Ln) (n\cdot D)^n N(Ln)=\int d\beta \, s_R(\beta) \, u(Ln)W_R[Ln,\beta n]N(\beta n),
\end{eqnarray}
where $s_{R}$ is the generating function for the LECs $c_{U_R}$. The object $W$ is the chiral analog of the Wilson line. It is defined as
\begin{eqnarray}\label{def:Wilson}
W_R[x_1,x_2]=P\exp\(\int_{x_1}^{x_2}dx^\mu A_\mu(x)\),\qquad W_L[x_1,x_2]=P\exp\(\int_{x_1}^{x_2}dx^\mu \widetilde{A}_\mu(x)\),
\end{eqnarray}
where the fields $A$ and $\widetilde{A}$ are defined in (\ref{def:A}) and depend on the parameter $a$.

The chiral Wilson lines $W_{R/L}$ obey the usual properties of gauge links for the non-linear realization of the chiral transformation (\ref{A->hAh+hdh}). Namely, they transform as a gauge link,
\begin{eqnarray}
W_{R}[x_1,x_2]\to h(x_1) W_{R}[x_1,x_2]h^\dagger(x_2),
\qquad
W_{L}[x_1,x_2]\to h(x_1) W_{L}[x_1,x_2]h^\dagger(x_2),
\end{eqnarray}
conjugation reverses the path of the link
\begin{eqnarray}
W_{R}[x_1,x_2]^\dagger =W_{R}[x_2,x_1],
\qquad
W_{L}[x_1,x_2]^\dagger =W_{L}[x_2,x_1],
\end{eqnarray}
and they satisfy the equations
\begin{eqnarray}\label{DW=0}
D_xW_{R}[x,z]=W_{R}[x,z]\overleftarrow{D}_z=0,\qquad \widetilde{D}_xW_{L}[x,z]=W_{L}[x,z]\overleftarrow{\widetilde{D}}_z=0,
\end{eqnarray}
where the derivatives $\overleftarrow{D}_\mu=\overleftarrow{\partial}_\mu-A_\mu$ are taken along the path. Under the parity and charge conjugations the chiral Wilson lines transform as
\begin{eqnarray}\label{descrete1}
&&W_R[x_1,x_2]\xrightarrow{P} W_L[\tilde x_1,\tilde x_2],\qquad W_L[x_1,x_2]\xrightarrow{P} W_R[\tilde x_1,\tilde x_2],
\\\label{descrete2}
&&W_R[x_1,x_2]\xrightarrow{C} \(W_L[x_2,x_1]\)^T,\qquad W_L[x_1,x_2]\xrightarrow{C} \(W_R[x_2,x_1]\)^T,
\end{eqnarray}
where $T$ indicates the matrix transposition. In the particular cases $a=\pm 1$, the fields (\ref{def:A}) take a pure-gauge form, and the Wilson lines $W_{R/L}$ become path-independent
\begin{eqnarray}\label{W-trivial}
W_{R}[x_1,x_2]\Big|_{a=+1}=W_{L}[x_1,x_2]\Big|_{a=-1}=u(x_1)u^\dagger(x_2),
\quad
W_{R}[x_1,x_2]\Big|_{a=-1}=W_{L}[x_1,x_2]\Big|_{a=+1}=u^\dagger(x_1)u(x_2).
\end{eqnarray}
In ref.~\cite{Feldmann:2026tex} analogous constructions were considered in the particular case $a=0$ (i.e. in the case $A_\mu=\Gamma_\mu$).

\subsection{Nucleon semi-compact operators}

Summing the tower of local operators (\ref{op:URn}, \ref{op:URl}), one gets semi-compact operators expressed via the Wilson lines $W_{R/L}$, which, in turn, can be used to construct TMD operators. For future convenience, we introduce the building blocks for such operators
\begin{eqnarray}\label{def:Psi}
\Psi_R(xn)=u(\infty n)W_R[\infty n,xn]N(xn),\qquad \Psi_L(xn)=u^\dagger(\infty n)W_L[\infty n,xn]N(xn),
\end{eqnarray}
and their conjugates
\begin{eqnarray}\label{def:Psi-bar}
\overline{\Psi}_R(xn)=\overline{N}(xn)W_R[xn,\infty n]u^\dagger(\infty n),\qquad \overline{\Psi}_L(xn)=\overline{N}(xn)W_L[xn,\infty n]u(\infty n).
\end{eqnarray}
Under the chiral transformation they transform as
\begin{eqnarray}
\Psi_R\to R \Psi_R,\qquad \Psi_L\to L \Psi_L,
\qquad
\overline{\Psi}_R\to \overline{\Psi}_R R^\dagger,\qquad \overline{\Psi}_L\to \overline{\Psi}_L L^\dagger.
\end{eqnarray}

The blocks $\Psi$ and $\overline{\Psi}$ are the analogs of the QCD semi-compact operators $U_1$ for the nucleon field\footnote{Note that one can also introduce another set of operators analogous to (\ref{def:Psi}, \ref{def:Psi-bar}, \ref{def:Pi}) with the same quantum numbers exchanging $a\to -a$ (or equivalently $W_R\leftrightarrow W_L$). These operators have the same transformation rules, and the same twist. These new operators should be accompanied by different LECs, and, in principle, added to the basis. Even worse, one can consider local operators (\ref{op:URn}, \ref{op:URl}) with various combinations of $D$ and $\widetilde{D}$, creating an infinite set of effective operators with the same quantum numbers. However, the TMD operators (constructed in the next sections) made of $\Psi$ and $\Pi$ and those made of these additional operators are equivalent up to terms $\sim \pi^3$ (at least). Therefore, up to the second chiral order one can absorb these extra operators into the constructed ones. These extra effects become important at higher orders of the chiral expansion, which is far beyond the scope of the present study, and hardly of practical interest.}. They have the same geometrical twist, as follows from the expansion (\ref{op:summing-U}), and thus are suitable leading chiral order candidates for the twist-1 semi-compact operators in ChPT. In this picture, the fields $u(\infty n)$ play the same role as the transverse Wilson lines in QCD. That is, they absorb the gauge transformation and guarantee the gauge invariance of the construction. 

The operators (\ref{def:Psi}, \ref{def:Psi-bar}) can also be seen as propagators of the nucleon field in the background of the strong force, from the point of creation to the point of measurement. Along this path, they preserve chirality, similarly to the quark fields in the QCD analogs. In this sense, they can be seen as the eikonalization of the nucleon propagator. This interpretation restricts the parameter $a=g_A$. However, we keep the parameter $a$ free until the sec.~\ref{sec:sf}, where it is fixed by the renormalization condition for rapidity divergences.

\subsection{Pion semi-compact operators and alternative form of the chiral Wilson line}

A similar construction can be made by summing (\ref{op:URn}, \ref{op:URl}) with a pion field instead of the nucleon one. The corresponding semi-compact operators have the form
\begin{eqnarray}\label{def:Pi}
\Pi_R(xn)=u(\infty n)W_R[\infty n,xn]u^\dagger(xn),\qquad \Pi_L(xn)=u^\dagger(\infty n)W_L[\infty n,xn]u(xn).
\end{eqnarray}
Their transformation properties are
\begin{eqnarray}\label{transform:pi}
\Pi_R\to R \Pi_R R^\dagger,\qquad \Pi_L\to L \Pi_L L^\dagger.
\end{eqnarray}
Together, the operators $\Pi$ and $\Psi$ constitute the complete leading-chiral-order set of semi-compact operators. At $a=0$, our blocks coincide (up to the factor $u(\infty n)$) with the blocks $\Xi$ (for $\Pi$) and $\mathcal{H}_v$ (for $\Psi$) used in ref.~\cite{Feldmann:2026tex}.

It is interesting to observe that these operators can be simplified by rewriting the Wilson lines in an alternative form. To do so, we note that
\begin{eqnarray}
A_\mu=u^\dagger\partial_\mu u + \frac{1+a}{2} u^\dagger (U \partial_\mu U^\dagger)u,
\qquad
\widetilde{A}_\mu=u\partial_\mu u^\dagger + \frac{1+a}{2} u (U^\dagger \partial_\mu U)u^\dagger.
\end{eqnarray}
These expressions can be seen as gauge transformations (\ref{A->hAh+hdh}) of the fields $(1+a)U \partial_\mu U^\dagger/2$ and $(1+a)U^\dagger \partial_\mu U/2$ with $h=u^\dagger$ and $h=u$, respectively. Thus, the Wilson lines $W_{R/L}$ can be written as
\begin{eqnarray}
W_R[x_1,x_2]=u^\dagger(x_1)\Pi_R[x_1,x_2]u(x_2),\qquad W_L[x_1,x_2]=u(x_1)\Pi_L[x_1,x_2]u^\dagger(x_2),
\end{eqnarray}
where
\begin{eqnarray}\label{def:Wilson-PI}
\Pi_R[x_1,x_2]=P\exp\(\frac{1+a}{2} \int_{x_1}^{x_2}dx^\mu U(x) \partial_\mu U^\dagger(x)\),\quad
\Pi_L[x_1,x_2]=P\exp\(\frac{1+a}{2} \int_{x_1}^{x_2}dx^\mu U^\dagger(x) \partial_\mu U(x)\).
\end{eqnarray}
At $x_1=\infty n$, the Wilson lines $\Pi_{R/L}$ turn into the semi-compact operators (\ref{def:Pi}), i.e., $\Pi_{R/L}[\infty n,xn]=\Pi_{R/L}(xn)$. The nucleon semi-compact operators (\ref{def:Psi}) written via $\Pi$ get rid of the factors $u(\infty n)$ and turn into
\begin{eqnarray}\label{def:Psi_V}
\Psi_R(xn)=\Pi_R(xn)u(xn)N(xn),\qquad \Psi_L(xn)=\Pi_L(xn)u^\dagger(xn)N(xn).
\end{eqnarray}

Note that although $U \partial_\mu U^\dagger$ is a pure-gauge term, the Wilson lines $\Pi_{R/L}$ do not trivialize, due to the non-unit factor $(1+a)/2$. In the cases $a=\pm 1$, the exponent vanishes or becomes a pure gauge, and the lines trivialize to (see also (\ref{W-trivial}))
\begin{eqnarray}
&&\Pi_R\Big|_{a=-1}=\Pi_L\Big|_{a=-1}= 1,
\qquad
\Pi_R[x_1,x_2]\Big|_{a=+1}=U(x_1) U^\dagger(x_2),\qquad \Pi_L[x_1,x_2]\Big|_{a=+1}=U^\dagger(x_1) U(x_2).
\end{eqnarray}
In these limits, the building blocks $\Psi$ reduce to
\begin{eqnarray}
&&\Psi_R\Big|_{a=-1}=uN,\qquad \Psi_L\Big|_{a=-1}=u^\dagger N,
\qquad
\Psi_R\Big|_{a=+1}=U(\infty n) u^\dagger N,\qquad \Psi_L\Big|_{a=+1}= U^\dagger(\infty n) uN.
\end{eqnarray}

Analogously to the nucleon operators, the operators (\ref{def:Pi}) can be seen as propagators of the pion field in the background of the strong force, preserving the chirality. This would correspond to the eikonalization of the pion propagator, and would fix the parameter $a=0$, such that the vertices $U\partial_\mu U^\dagger$ reproduce the $4\pi$ interaction with the pion background. This is confirmed later, in sec.~\ref{sec:sf}, by the renormalization condition for rapidity divergences.

\subsection{On generating functions for LECs}

At the next stage, the semi-compact operators can be composed into the TMD operators following the general structure (\ref{semi-compact->TMD}). Before proceeding to this construction, let us comment on the appearance of the LEC generating function. As shown in (\ref{op:summing-U}), the semi-compact operator at point $x$ is obtained from the blocks $\Psi$ (or $\Pi$) by integrating them with a function $s(\beta)$, $U(x)\sim \int d\beta\, s(\beta)\Psi(\beta x)$. Combining two such operators, one gets a structure of the form
\begin{eqnarray}
\overline{U}_1(z_1 n+b)\dots U_1(z_2 n)\sim \int d\beta_1 d\beta_2 \, s_1(\beta_1)s_2(\beta_2)\, \overline{\Psi}(\beta_1 z_1n+b) \dots\Psi(\beta_2 z_2n).
\end{eqnarray}
Forward matrix elements are invariant under a total shift of the operator, and thus one can reduce the independent dependence on $\beta_1$ and $\beta_2$ to a single-variable dependence, obtaining an expression of the form
\begin{eqnarray}
\overline{U}_1(z n+b)\dots U_1(0)\sim \int d\beta \, F(\beta)\, \overline{\Psi}(\beta z n+b) \dots\Psi(0),
\end{eqnarray}
where $F(\beta)\sim \int d\beta_2 s_1(\beta+\beta_2)s_2(\beta_2)$. Furthermore, an accurate construction of the effective operator also requires an expansion in $b$ (which is the usual Taylor expansion). Each of these operators has its own LEC, and thus, after the summation of this expansion, the generating functions $F$ acquire a dependence on $b$. Therefore, in what follows, the TMD operators are constructed from the building blocks $\Psi$ and $\Pi$ positioned at $\beta zn+b$ and $0$, and accompanied by a generating function of LECs $F(\beta,b)$. In addition, the function $F(\beta,b)$ can be multiplied by tensor structures built from the vector $b^\mu$ if such a construction is required.

\section{TMD operators in ChPT (nucleon case)}
\label{sec:nucleon}

In this section, we compose the TMD-twist-(1,1) operators $O^{[\Gamma]}$ at the leading chiral order using nucleon building blocks, and find the leading value for corresponding generating functions. We start from this case, because its almost identical to the QCD form, despite many independent components that appear in the final expression. 

The TMD operators with $\gamma^+_R=\gamma^+(1+\gamma^5)/2$ and $\gamma^+_L=\gamma^+(1-\gamma^5)/2$ are chiral-even and transform under the right and left chiral transformations of quarks, respectively. Consequently, their chiral analogs can be constructed as $\overline{\Psi}_R\dots\Psi_R$ and $\overline{\Psi}_L\dots\Psi_L$, respectively. The TMD operator with $i\sigma^{\alpha+}\gamma^5$ is chiral-odd, and thus it is expressed via $\overline{\Psi}_R\dots\Psi_L$ and $\overline{\Psi}_L\dots\Psi_R$. So, we introduce
\begin{eqnarray}
O_N^{[\gamma^+_R],A}(z,b)&=&\int d\beta \overline{\Psi}_R(\beta\, zn+b)F^R_\Psi\frac{\tau^A}{2}\Psi_R(0),
\\
O_N^{[\gamma^+_L],A}(z,b)&=&\int d\beta \overline{\Psi}_L(\beta\, zn+b)F^L_\Psi\frac{\tau^A}{2}\Psi_L(0),
\\
O_N^{[i\sigma^{\alpha+}\gamma^5],A}(z,b)&=&\int d\beta \(
\overline{\Psi}_L(\beta\, zn+b) F^{\alpha,RL}_{\Psi}\frac{\tau^A}{2}\Psi_R(0)
+
\overline{\Psi}_R(\beta\, zn+b) F^{\alpha,LR}_{\Psi}\frac{\tau^A}{2}\Psi_L(0)\),
\end{eqnarray}
where the $F$'s are generating functions of $(\beta,b)$, which also carry spinor indices, and $\tau^A$ couples the flavor indices. 

It is important that the transformation rules refer to \textit{quarks}, not to \textit{nucleons}, whose spin structure is entirely independent. Therefore, the nucleon spinors can be coupled by any Dirac matrix that does not violate the transformation properties of the operator. Therefore, the functions $F_\Psi$ can have any Dirac structure, not only those that appear in the QCD case.

The matrices $F_\Psi$ can be built from the standard elements of the Dirac basis, and should contain one $\gamma^+$, with the remaining indices forming a boost-invariant combination, in order to match the Lorentz transformation properties of the original QCD operators. Therefore, the complete decomposition of $F_\Psi$ reads
\begin{eqnarray}\label{F-decomp1}
F_\Psi^{R}&=& F_1^R\gamma^++G_1^R\gamma^+\gamma^5+iF_{1T}^R \, \epsilon^{\mu\nu}_T b_\mu i\sigma^{\nu+} \gamma^5+iG_{1T}^R b_\nu i\sigma^{\nu+} \gamma^5,
\\\label{F-decomp2}
F_\Psi^{\alpha,RL}&=& H_1^{RL}i\sigma^{\alpha+}\gamma^5+H_{1TT}^{RL}\(b^\alpha b_\mu -\frac{1}{2}\delta^{\alpha}_\mu b^2\)i\sigma^{\mu+}\gamma^5
+H_{1L}^{RL}\,ib^\alpha \gamma^+\gamma^5
+H_{1T}^{RL}\,i\epsilon_T^{\alpha \mu}b_\mu \gamma^+
\\\nn &&
+\Delta H_1^{RL}i\epsilon_T^{\alpha\mu}i\sigma^{\mu +}\gamma^5+\Delta H_{1TT}^{RL}\(i\epsilon_T^{\alpha\mu}b_\mu b_\nu -\frac{1}{2}i\epsilon_T^{\alpha\nu} b^2\)i\sigma^{\nu+}\gamma^5
+\Delta H_{1L}^{RL}\,\epsilon_T^{\alpha\mu}b^\mu \gamma^+\gamma^5
+\Delta H_{1T}^{RL}\,b^\alpha \gamma^+
,
\end{eqnarray}
where the generating functions are functions of $(\beta, b)$. The analogous decomposition holds for $F_\Psi^{L}$ and $F_\Psi^{\alpha,LR}$ with the corresponding replacement of indices. In these expansions, we have chosen the factors $i$ to conveniently match the discrete symmetry transformations.

Further restrictions on the effective operators come from the discrete symmetries, namely, hermiticity, parity, charge conjugation and time reversal. The transformation rules for the chiral Wilson lines are given in (\ref{descrete1}, \ref{descrete2}), and those for the other fields can be found in the literature (see, for instance, ref.~\cite{Scherer:2012xha} for the transformation rules of chiral fields, and refs.~\cite{Rodini:2022wki, Boer:2003cm} for the transformation of TMD operators). In particular, the parity transformation requires the following relations between the generating functions \cite{Moiseeva:2012zi}
\begin{eqnarray} 
F^L=F^R,\qquad G^L=-G^R,\qquad H^{RL}=H^{LR}, \qquad \Delta H^{RL}=-\Delta H^{LR},
\end{eqnarray}
for all lower indices. Meanwhile, the PT transformation implies that the functions $F_{1T}$, $H_{1T}$ and $\Delta H_{1T}$ are T-odd, i.e., they change sign under $L\to-L$. This corresponds to the sign-change relation between the Drell-Yan and SIDIS definitions of the Sivers and Boer-Mulders functions \cite{Collins:2002kn}.

Summarizing these observations, we obtain the following expressions for the effective TMD operators
\begin{eqnarray}\label{op:+}
O_N^{[\gamma^+],A}(z,b)&=&\int d\beta\Big[
F_1 \(O_{RR}^{[\gamma^+]}+O_{LL}^{[\gamma^+]}\)
+G_1 \(O_{RR}^{[\gamma^+\gamma^5]}-O_{LL}^{[\gamma^+\gamma^5]}\)
\\\nn &&
\quad
+i\epsilon_T^{\mu\nu}b_\mu F_{1T} \(O_{RR}^{[i\sigma^{\nu+}\gamma^5]}+O_{LL}^{[i\sigma^{\nu+}\gamma^5]}\)
+ib_\mu G_{1T} \(O_{RR}^{[i\sigma^{\mu+}\gamma^5]}-O_{LL}^{[i\sigma^{\mu+}\gamma^5]}\)
\Big],
\\\label{op:+5}
O_N^{[\gamma^+\gamma^5],A}(z,b)&=&\int d\beta\Big[
F_1 \(O_{RR}^{[\gamma^+]}-O_{LL}^{[\gamma^+]}\)
+G_1 \(O_{RR}^{[\gamma^+\gamma^5]}+O_{LL}^{[\gamma^+\gamma^5]}\)
\\\nn &&
\quad
+i\epsilon_T^{\mu\nu}b_\mu F_{1T} \(O_{RR}^{[i\sigma^{\nu+}\gamma^5]}-O_{LL}^{[i\sigma^{\nu+}\gamma^5]}\)
+ib_\mu G_{1T} \(O_{RR}^{[i\sigma^{\mu+}\gamma^5]}+O_{LL}^{[i\sigma^{\mu+}\gamma^5]}\)
\Big],
\\\label{op:a+5}
O_N^{[i\sigma^{\alpha +}\gamma^5],A}(z,b)&=&\int d\beta\Big[
\(H_1 \delta^\alpha_\mu+H_{1TT}\(b^\alpha b_\mu -\frac{1}{2}\delta^{\alpha}_\mu b^2\)\) \(O_{RL}^{[i\sigma^{\mu+}\gamma^5]}+O_{LR}^{[i\sigma^{\mu+}\gamma^5]}\)
\\\nn &&\quad
+i \epsilon_T^{\alpha \mu}b_\mu H_{1T}\(O_{RL}^{[\gamma^+]}+O_{LR}^{[\gamma^+]}\)
+i b^\alpha H_{1L}\(O_{RL}^{[\gamma^+\gamma^5]}+O_{LR}^{[\gamma^+\gamma^5]}\)
\\\nn &&
+\(\Delta H_1i\epsilon^{\alpha\mu}_T+\Delta H_{1TT}\(i\epsilon^{\alpha\nu}_T b_\nu b_\mu -\frac{i}{2}\epsilon^{\alpha\mu}_Tb^2\)\) \(O_{RL}^{[i\sigma^{\mu+}\gamma^5]}-O_{LR}^{[i\sigma^{\mu+}\gamma^5]}\)
\\\nn &&\quad
+b^\alpha \Delta  H_{1T}\(O_{RL}^{[\gamma^+]}-O_{LR}^{[\gamma^+]}\)
+\epsilon^{\alpha\mu}_Tb_\mu \Delta H_{1L}\(O_{RL}^{[\gamma^+\gamma^5]}-O_{LR}^{[\gamma^+\gamma^5]}\)
\Big],
\end{eqnarray}
where
\begin{eqnarray}
O_{XY}^{[\Gamma]}=\overline{\Psi}_X\Gamma\frac{\tau^A}{2}\Psi_Y.
\end{eqnarray}
Therefore, in total we observe twelve independent generating functions. The vector and axial distributions match the ordinary decomposition. Interestingly, these operators depend on the same set of four generating functions, which implies some mutual dependence between the helicity and unpolarized sectors. In QCD, such a dependence is observed only between the Sivers and worm-gear functions, which are expressed through the same collinear distribution of twist three \cite{Scimemi:2018mmi, Rein:2022odl}. The chiral-odd sector has eight independent structures; however, four of them can appear only due to chirality mixing, as we see below.

The expressions for the operators (\ref{op:+}, \ref{op:+5}, \ref{op:a+5}) exactly match the decomposition of TMD distributions (\ref{def:+}, \ref{def:+5}, \ref{def:a+5}). To see this, we compute the forward matrix elements of these operators (\ref{def:op->TMD}) at tree level. At this order, the Dirac structure is sandwiched between nucleon spinors, and after applying the standard relations $\bar u \gamma^+ u=2p^+$, $\bar u \gamma^+\gamma^5u=2p_+s_L$, and $\bar u i\sigma^{\alpha +}\gamma^5 u=2p_+ s_T^\alpha$, one finds the correspondence
\begin{align}
&F_1(x,b)=\mathring{f}_1(x,b), &&F_{1T}(x,b)=M \mathring{f}_{1T}^\perp(x,b),
&&G_1(x,b)=\mathring{g}_1(x,b), &&G_{1T}(x,b)=M \mathring{g}_{1T}^\perp(x,b),
\\
&H_1(x,b)=\mathring{h}_1(x,b), &&H_{1TT}(x,b)=\frac{M^2}{2}\mathring{h}_{1T}^\perp(x,b),
&&H_{1T}(x,b)=M \mathring{h}_1^\perp(x,b), &&H_{1L}(x,b)=-M\mathring{h}_{1L}^\perp(x,b),
\end{align}
where $\mathring{f}_1$ denotes the unpolarized TMDPDF in the chiral limit, and analogously for the other TMDPDFs. These functions are independent for each isospin representation, and they are related to the flavor indices of the TMDPDFs as 
\begin{eqnarray}
\mathring{f}_{fg/N}=2\Big(\delta_{fg}F^{I=0}+\sum_{a=1}^{3}\langle N|\tau^a|N\rangle\,\tau^a_{gf}\,F^{I=1}\Big),
\end{eqnarray}
where $I$ is the isospin representation ($I=0$ isoscalar, $I=1$ isovector). In the case of proton it gives
\begin{eqnarray}
\mathring{f}_{u/p}=2(F^{I=0}+F^{I=1}),\qquad \mathring{f}_{d/p}=2(F^{I=0}-F^{I=1}).
\end{eqnarray}
The anti-quark distributions are obtained at negative values of $x$ by usual rules \cite{Boer:2003cm, Scimemi:2018mmi}. The generating functions $\Delta H$ do not contribute at leading order. They first contribute at one loop, accompanied by the coupling constant $g_A$.

At the $b=0$ limit, the Wilson lines of the semi-compact operators vector and axial operators cancel, leading to the conventional structure for effective chiral operator with two functions $F_1$ and $G_1$. This structure (at $a=-1$) has been discussed \cite{Moiseeva:2012zi}. Note, that for the case of chiral-odd operator, Wilson lines do not cancel except the spacial case of $a=0$. This matches with observation made in ref.~\cite{Diehl:2005rn} that construction of chiral-odd operators requires extra insertions of $u_\mu$.

\section{TMD operators in ChPT (pion case)}
\label{sec:pion}

In this section, we compose the TMD-twist-(1,1) operators $O^{[\Gamma]}$ at the leading chiral order using pion building blocks. This construction differs from the nucleon one, because the pion does not carry a spin structure, and thus additional Lorentz indices should be obtained in a different manner.

To construct the vector operator from pion fields, one should insert an additional derivative acting between the semi-compact operators \cite{Kivel:2002ia}, $\Pi^\dagger (\overrightarrow{\partial}_\mu-\overleftarrow{\partial}_\mu)\Pi$. Contracted with $n^\mu$, this operator acquires the required properties. Let us introduce the additional building blocks
\begin{eqnarray}\label{def:X}
X_R(xn)&=&u(\infty n)W_R[\infty n,xn](n\cdot u)u^\dagger(xn)=-i\Pi_R(xn) U \partial_+ U^\dagger(xn),
\\
X_L(xn)&=&u^\dagger(\infty n)W_L[\infty n,xn](n\cdot u)u(xn)=i\Pi_L(xn) U^\dagger \partial_+ U(xn).
\end{eqnarray}
These elements correspond to derivatives of the field $\Pi$, since due to (\ref{DW=0})
\begin{eqnarray}
(n\cdot \partial)\Pi_R=\frac{i(1+a)}{2}X_R,\qquad
(n\cdot \partial)\Pi_L=-\frac{i(1+a)}{2}X_L.
\end{eqnarray}
Using these fields, one can construct the analogs of the right and left vector operators as
\begin{eqnarray}
O^{[\gamma^+_R]}_\pi(z,b)&=& \int d\beta F_{1\pi}^R(\beta,b)\(\Pi^\dagger_R(\beta z n+b) X_R(0)+X^\dagger_R(\beta z n+b) \Pi_R(0)\),
\\
O^{[\gamma^+_L]}_\pi(z,b)&=& \int d\beta F_{1\pi}^L(\beta,b)\(\Pi^\dagger_L(\beta z n+b) X_L(0)+X^\dagger_L(\beta z n+b) \Pi_L(0)\).
\end{eqnarray}
Here, the isospin indices of the pion fields define the flavor content of the operator. Note that the generating functions $F_{1\pi}^{R,L}$ are scalars, and, in contrast to the nucleon case (\ref{F-decomp1}, \ref{F-decomp2}), cannot accommodate more degrees of freedom. 

The parity transformation sets $F_{1\pi}^R=-F_{1\pi}^L$, which leads to the definition of a unique vector TMD operator
\begin{eqnarray}\label{pion:vec}
O^{[\gamma^+]}_\pi(z,b)&=& \frac{F_\pi^2}{4}\int d\beta F_{1\pi}(\beta,b)\(\Pi^\dagger_R X_R+X^\dagger_R \Pi_R
-
\Pi^\dagger_L X_L-X^\dagger_L \Pi_L
\),
\end{eqnarray}
where all pairs of fields have arguments as in $\Pi^\dagger_R(\beta z n+b)X_R(0)$. The factor $F_\pi^2/4$ is introduced for future convenience. In the limit $a=1$ and $b=0$, this operator coincides with the operator defined in ref.~\cite{Kivel:2002ia}, which describes the pion PDF and GPD.

Let us mention that we omit the construction of the pion helicity operator $O^{[\gamma^+\gamma^5]}\sim O^{[\gamma^+_R]}-O^{[\gamma^+_L]}$. The reason is that its matrix elements vanish in pure-pion ChPT. Nonetheless, it contributes to the nucleon matrix elements beyond leading order, representing the contribution of the pion cloud to the nucleon helicity structure. Specifically, the leading term of this operator that is not a total derivative is $\sim \pi^3$, so it contributes starting from two loops at the earliest. Since we are not interested in higher-order chiral effects, we omit this operator.

\begin{figure}[t]
\begin{center}
\includegraphics[width=0.86\textwidth]{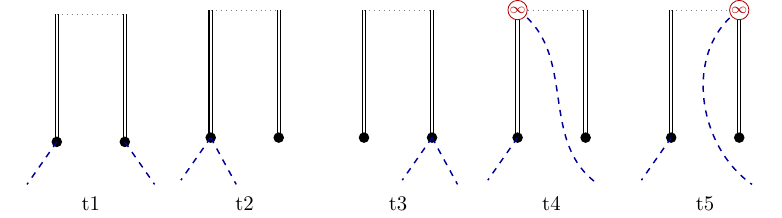}
\end{center}
\caption{\label{fig:tree} Tree-level diagrams contributing to the pion unpolarized distribution. The mirror-symmetric diagrams for t4 and t5 are not shown.}
\end{figure}

To find the interpretation of the generating function $F_{1\pi}$, we compute the matrix elements of these operators at tree level. In contrast to the nucleon case, where there is a single diagram and the whole computation consists of the evaluation of bispinor products, in the pion case the leading order contains several terms with a richer structure. To reveal them, the operator (\ref{pion:vec}) should be expanded up to $\pi^2$ terms. We obtain
\begin{eqnarray}\label{pion:vec-pi2}
O^{[\gamma^+]}_{\pi}(z,b)&=&\frac{-i}{4}\int d\beta F_{1\pi}(\beta,b)\Big[(1+a)\hat \pi(y)\overleftrightarrow{\partial_+}\hat \pi(0)
\\\nn &&
\qquad
+i(1-a)\epsilon^{abc}\tau^c\(\pi^a(y) \partial_+ \pi^b(y)+\pi^a(0) \partial_+ \pi^b(0)\)
\\\nn && 
\qquad
-(1+a)\(\Delta \hat \pi_\infty \partial_+ \hat \pi(0)+\partial_+ \hat \pi(y)\Delta \hat \pi_\infty\)+\mathcal{O}(\pi^4)\Big],
\end{eqnarray}
where the argument $y$ stands for $\beta z n+b$, $\hat \pi=\pi^a \tau^a$, $\Delta \hat \pi_\infty=\hat \pi(\pm \infty n+b)-\hat \pi(\pm \infty n)$, and we have dropped the total-derivative terms in (\ref{pion:vec-pi2}), since they do not contribute to forward matrix elements. A novel feature is the appearance of pion fields at the light-cone infinity (see diagrams t4 and t5 in fig.~\ref{fig:tree}). At tree level, these terms vanish, because they imply the propagation of a (massive) pion to infinity, which produces a damping phase factor $\sim e^{\pm iL p^+}$. 

At tree level, the pion matrix element receives contributions from three non-vanishing diagrams, t1, t2 and t3 in fig.~\ref{fig:tree}. Their computation gives
\begin{eqnarray}\label{pi:tree-1}
\langle \pi^b(p)|O_\pi^{[\gamma^+]}(z,b)|\pi^a(p)\rangle&=& 
p_+\int d\beta F_{1\pi}(\beta,b)\Big[i(1-a)\epsilon^{abc}\tau^c +\frac{1+a}{2}\(\tau^a\tau^b e^{-i\beta zp_+}-\tau^b\tau^a e^{i\beta zp_+}\)\Big],
\end{eqnarray}
where the second term comes from the two possibilities to attach the external pion fields. Following ref.~\cite{Kivel:2002ia}, we define the pion TMDPDFs with given isospin
\begin{eqnarray}\label{pi:I=0}
&&\int \frac{dz}{2\pi}e^{-ixp^+z} \langle \pi^b(p)|\tr[O_\pi^{[\gamma^+]}(z,b)]|\pi^a(p)\rangle =2\delta^{ab}f_{1\pi}^{I=0}(x,b),
\\\label{pi:I=1}
&&\int \frac{dz}{2\pi}e^{-ixp^+z} \langle \pi^b(p)|\tr[\tau^c O_\pi^{[\gamma^+]}(z,b)]|\pi^a(p)\rangle =4i\epsilon^{abc}f_{1\pi}^{I=1}(x,b).
\end{eqnarray}
Inserting (\ref{pi:tree-1}), we obtain
\begin{eqnarray}\label{pi:tree-4}
\mathring{f}_{1\pi}^{I=0}(x,b)&=&\frac{(1+a)}{2}\(F_{1\pi}(-x,b)-F_{1\pi}(x,b)\),
\\\label{pi:tree-5}
\mathring{f}_{1\pi}^{I=1}(x,b)&=&\frac{(1+a)}{2}\frac{F_{1\pi}(-x,b)+F_{1\pi}(x,b)}{2}+\frac{(1-a)}{2}\delta(x)\int d\beta F_{1\pi}(\beta,b).
\end{eqnarray}
For $x\neq 0$, these relations identify the antisymmetric and symmetric combinations of $F_{1\pi}$ as the isoscalar and isovector pion TMDPDFs in the chiral limit. At $a=1$, this expression coincides with the one derived in ref.~\cite{Kivel:2002ia}. The last term in (\ref{pi:tree-5}) represents the contact term, which is not visible at $x\neq 0$. In the GPD case, such terms are responsible for the generation of the D-term, and they also appear in ref.~\cite{Kivel:2002ia}. In some sense, they could be omitted, but it is known that higher chiral loops produce higher derivatives of the delta function, which could sum up to a contribution that does not vanish at $x\neq0$ (see refs.~\cite{Kivel:2008ry, Kivel:2008mf, Moiseeva:2012zi}).

The operator (\ref{pion:vec}) corresponds to the unpolarized TMDPDF of the pion. However, the pion also has the Boer-Mulders contribution $h_1^\perp$ (\ref{def:a+5}). Such a contribution corresponds to a chiral-odd operator that transforms as $R\dots L^\dagger$ or $L\dots R^\dagger$. However, the combinations $\Pi_R^\dagger \Pi_L$ and $\Pi_L^\dagger \Pi_R$ do not transform correctly, because the transformation of the $u$ fields at infinity (see (\ref{transform:pi})) is not compensated, and leads to a product $R^\dagger L$ or $L^\dagger R$. In order to compensate it, one should introduce an extra factor $U$, which turns $R\leftrightarrow L$ (\ref{sym:chiral}). It can be placed at either of the infinities, $(Ln+b)$ or $(Ln)$. To take this into account, we introduce the factor
\begin{eqnarray}
\Delta U_\infty = U(\pm \infty n+b)-U(\pm \infty n).
\end{eqnarray}
Then the chiral-odd operator, which corresponds to the Boer-Mulders contribution in the pion, takes the form
\begin{eqnarray}\label{pion:tensor}
O^{[i\sigma^{\alpha+}\gamma^5]}_\pi(z,b)&=& \frac{F_\pi^2}{4}\int d\beta \epsilon^{\alpha \mu}_T b_\mu H_{1T\pi}(\beta,b)\(\Pi^\dagger_R \Delta U_\infty X_L-X^\dagger_R \Delta  U_\infty\Pi_L
-
\Pi^\dagger_L\Delta U^\dagger_\infty X_R+X^\dagger_L \Delta U^\dagger_\infty \Pi_R
\),
\end{eqnarray}
where we extracted the index $\alpha$ from the generating function and placed it in the appropriate form.

The expansion of the operator (\ref{pion:tensor}) gives
\begin{eqnarray}
\label{pion:tensor-pi2}
O^{[i\sigma^{\alpha+}\gamma^5]}_{\pi}(z,b)&=&\frac{-i\epsilon^{\alpha\mu}_T b_\mu}{2}\int d\beta H_{1T\pi}(\beta,b)
\Big[\Delta \hat \pi_\infty \partial_+ \hat \pi(0)-\partial_+ \hat \pi(y)\Delta \hat \pi_\infty+\mathcal{O}(\pi^4)\Big].
\end{eqnarray}
This expression is entirely proportional to $\Delta \pi_\infty$, which is the result of the fact that $\Delta U_\infty \sim \Delta \pi_\infty+\mathcal{O}(\pi^2)$. Therefore, at tree level, the Boer-Mulders function receives contributions only from diagrams of type t4 and t5, and thus vanishes. Note that the matrix element does not vanish at higher chiral orders (see sec.~\ref{sec:1loop}), because the interaction with infinity within a loop can produce a finite expression.

It should be noted that the insertion of $\Delta U_\infty$ is not the unique way to change the chiral transformation properties. One can also insert $U$ at the end-points of the Wilson lines (or, generally, at any point of the Wilson line, but this would violate boost invariance). Such a definition leads to a different operator, which also vanishes at tree level, although for a different reason (this operator turns out to be a total derivative at order $\pi^2$). So, in both cases, the main conclusion that follows from the construction of Boer-Mulders-type operators in the pure pion sector is that they are suppressed in chiral perturbation theory by at least one order.

\section{Chiral loop corrections and rapidity divergences}
\label{sec:loop}

The formulation of TMD operators in ChPT allows one to study the chiral corrections that arise through loops of pions and nucleons. However, the results of these computations are not very practical, since the values of the generating functions are not known, and at the next chiral order they should be accompanied by sub-leading chiral TMD operators with their own generating functions. Still, chiral loops contain interesting information, namely, the rapidity divergences. The presence of rapidity divergences is one of the defining properties of TMD operators, and it leads to the definition of the CS kernel. In this section, we demonstrate the presence of rapidity divergences in the TMD operators (\ref{op:+}, \ref{op:+5}, \ref{op:a+5}, \ref{pion:vec}), and explicitly compute the leading rapidity divergence for the nucleon and pion TMD operators. 

\subsection{One-loop diagrams}
\label{sec:1loop}

Rapidity divergences appear in the interaction of the fields in the bulk with the fields of the Wilson line located close to the light-cone infinity \cite{Vladimirov:2017ksc}. In QCD the expansion of the Wilson line has the form
\begin{eqnarray}
[\infty n,0]=1-ig\int_\infty^0 d\sigma A_+(\sigma)+\mathcal{O}(g^2).
\end{eqnarray}
The contraction with $A_+(\sigma)$ leads to the propagator $\sim (-2\sigma x^++x^2)^{-1}$, which generates a divergence at $\sigma\to\infty$. It is important that the propagation takes place at finite distances $x^2\sim b^2$, because for propagation at zero distance the resulting divergences are collinear and cancel in the TMD operator. It is also important that the point $x$ in this propagator is integrated over, because if it is a fixed point, it serves as a regulator and the integral is finite.

In ChPT, rapidity divergences appear by the same mechanism. The expansion of the Wilson line $\Pi_R$ (\ref{def:Wilson-PI}) reads
\begin{eqnarray}\label{W:expansion}
\Pi_R[x_1,x_2]=1+\frac{i(1+a)}{2F_\pi}\(\hat \pi(x_1)-\hat \pi(x_2)\)-\frac{(1+a)^2}{8F^2_\pi}\(\hat \pi^2(x_1)-2\hat \pi(x_1)\hat \pi(x_2)+\hat \pi^2(x_2)\)
\\\nn +\frac{1-a^2}{4F_\pi^2}i\epsilon^{abc}\tau^c\int_{x_1}^{x_2} d\sigma\, \pi^a(\sigma)\partial_+ \pi^b(\sigma)+\mathcal{O}(\pi^3),
\end{eqnarray}
where $\hat \pi=\pi^a \tau^a$. To present this expression in the current form, we have used integration by parts. The expression for $\Pi_L$ differs by the sign of the second term. Clearly, the integral in the second line is a potential source of rapidity divergences. In momentum space, it gives rise to the Feynman rule
\begin{eqnarray}
V^{ab}_W(k_1,k_2)&=&i\epsilon^{abc}\tau^c \frac{1-a^2}{4F_\pi^2}\frac{k_2^+-k_1^+}{k_1^++k_2^+\mp i0},
\end{eqnarray}
where the momenta $k_{1,2}$ are incoming to the vertex, and the sign of $\mp i0$ corresponds to the direction of the Wilson line, $\pm\infty n$. Note that the vertex is antisymmetric with respect to $k_1\leftrightarrow k_2$. This property eliminates many terms in the diagrams at the level of algebra.

Another potential source is the pion fields at the end-points of the Wilson lines. These terms appear from the operator itself, and also from the Wilson line (the first line of (\ref{W:expansion})). To accurately process such diagrams, one should regularize the infinity by a finite $L$, compute the diagram, and then take the limit $L\to\pm \infty$. Since a finite length of the Wilson line regulates rapidity divergences, such a divergence would reveal itself as $\ln(L)$.

\begin{figure}[t]
\begin{center}
\includegraphics[width=0.96\textwidth]{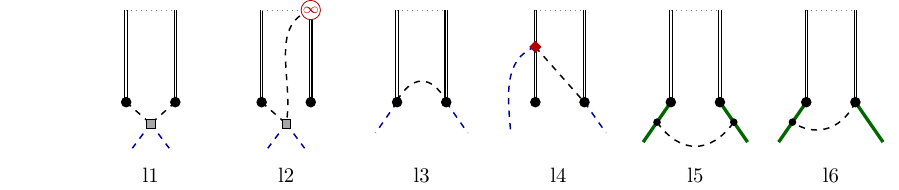}
\end{center}
\caption{\label{fig:1loop} Examples of one-loop diagrams. Double lines show Wilson lines, dashed lines show pion propagators, thick lines show nucleon propagators.}
\end{figure}

Due to the cumbersome structure of the operator, there are many diagrams already at one loop. Some of them are shown in fig.~\ref{fig:1loop}. The majority of diagrams are obviously rapidity-finite, such as the diagrams l1, l3, l5, l6 in fig.~\ref{fig:1loop}. However, there are also candidates for being divergent, such as the diagram l4, which contains an integral along the Wilson line, and l2, which contains the interaction with the pion at infinity.  We have explicitly computed all such diagrams and found that each of them is rapidity-finite. In many cases, it is obvious; for example, in the diagram l4 one end of the propagator is integrated over the line, but the other end is fixed, and thus regularizes the divergence. The diagrams of type l2 appear to be rapidity-finite but non-zero. This is an important observation, because it implies that the Boer-Mulders contribution to the pion, which vanishes at tree level, is non-zero at one loop.

Thus, the one-loop diagrams have no rapidity divergences. 

\subsection{Two-loop diagrams}

\begin{figure}[t]
\begin{center}
\includegraphics[width=0.86\textwidth]{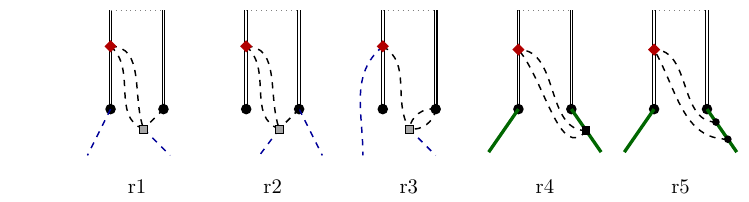}
\end{center}
\caption{\label{fig:2loop} The diagrams that produce the leading rapidity divergence. Mirror diagrams should be added.}
\end{figure}

A similar analysis performed at two loops shows that there are diagrams with rapidity divergences. They\footnote{We did not analyze all diagrams with one of the pion fields attached to infinity, assuming that they are rapidity-finite, as in the one-loop case.} are shown in fig.~\ref{fig:2loop}. In all these diagrams, the rapidity-divergent part appears through an integral of the form
\begin{eqnarray}
I(b)&=&-i\int \frac{d^dl}{(2\pi)^{d}}\frac{e^{i(l\cdot b)} B(l^2)}{(l^+\pm i\delta)[(l-p)^2-m^2_\pi+i0]}.
\end{eqnarray}
Here, $\delta$ is the parameter of the $\delta$-regularization \cite{Echevarria:2015byo, Echevarria:2016scs}. The function $B$ is the bubble integral
\begin{eqnarray}\label{def:B}
B(l^2)=\frac{2}{(4\pi)^{d/2}}\Gamma\(1-\frac{d}{2}\)\int_0^1 dx (m_\pi^2-x(1-x)l^2-i0)^{\frac{d}{2}-1},
\end{eqnarray}
with $x$ being the Feynman parameter. The rapidity-divergent part of the integral $I$ reads
\begin{eqnarray}
I(b)&=&\frac{-m_\pi^4}{(2\pi)^4 p^+} \mathcal{K}(b^2)\ln\frac{p^+}{\delta^+}+\text{(finite terms)},
\end{eqnarray}
where
\begin{eqnarray}\label{def:K}
\mathcal{K}(b^2)=\frac{4\pi^3}{ m_\pi^4}\int \frac{d^{d-2}l_T}{(2\pi)^{d-2}}e^{i(l\cdot b)}B(-l_T^2)=\frac{1}{m_\pi^2 b^2}\int_0^1 dx K_2\(\sqrt{\frac{m^2_\pi b^2}{x(1-x)}}\),
\end{eqnarray}
with $K_2$ being the Macdonald function. The factor $4\pi^3/ m_\pi^4$ is extracted for future convenience.

In the pion case there are three principal diagrams, shown as r1, r2 and r3 in fig.~\ref{fig:2loop}. There are also the corresponding mirror-conjugated diagrams, which are trivially obtained from the original ones. We have found that the rapidity-divergent parts of these diagrams are proportional to the tree-level diagrams with the same coefficient. Explicitly,
\begin{eqnarray}
\text{diag}_{r1/r2/r3}&=&8(1-a^2)\frac{m_\pi^4}{(4\pi F_\pi)^4}\ln\(\frac{p^+}{\delta^+}\) \mathcal{K}(b^2)\times \text{diag}_{t1/t3/t3}+\text{(finite terms)}.
\end{eqnarray}
Adding the mirror-conjugated diagrams, we obtain that the rapidity-divergent part is proportional to the tree-level expression (\ref{pi:tree-1})
\begin{eqnarray}\label{rap-div-pion}
\langle \pi^b|O_\pi^{[\gamma^+]}(z,b)|\pi^a\rangle\Big|_{\text{NNLO}}=
16(1-a^2)\frac{m_\pi^4}{(4\pi F_\pi)^4}\ln\(\frac{p^+}{\delta^+}\)\mathcal{K}(b^2)\times\langle \pi^b|O_\pi^{[\gamma^+]}(z,b)|\pi^a\rangle\Big|_{\text{tree}}+\text{(finite terms)}.
\end{eqnarray}
Note that the tree-level structure is reproduced entirely, together with the contact terms $\sim \delta(x)$. 

The expression (\ref{rap-div-pion}) is the same for both isospin channels. However, this is not a general rule, but a consequence of the simple structure of the leading diagrams. The isospin independence of the rapidity divergence of the pion operator is expected to be violated already at the next chiral order.

In the nucleon case, there are two diagrams, shown as r4 and r5 in fig.~\ref{fig:2loop}. It is simple to observe that the rapidity-divergent parts of these diagrams coincide (up to the coupling-constant factor $-g_A^2$ in the diagram r5). Also, they directly reproduce the tree-level matrix element and are independent of the Dirac structure. However, they depend on the isospin channel of the operator, because the $\tau^A$ matrix is positioned between the Wilson lines, and thus the group factor of the diagrams is
\begin{eqnarray}\label{def:cI}
\tau^a \tau^A \tau^a=c_I \tau^A,
\end{eqnarray}
with $c_{I=0}=3$ (i.e. for $\tau^A=\mathbb{I}$) and $c_{I=1}=-1$ (i.e. for $\tau^A=\tau^b$). Summing the diagrams, we obtain
\begin{eqnarray}\label{rap-div-nuc}
\langle N|O_N^{[\Gamma],A}(z,b)|N\rangle\Big|_{\text{NNLO}}=
4c_I(1-a^2)(1-g_A^2)\frac{m_\pi^4}{(4\pi F_\pi)^4}\ln\(\frac{p^+}{\delta^+}\)\mathcal{K}(b^2)\times\langle N|O_N^{[\Gamma],A}(z,b)|N\rangle\Big|_{\text{tree}}
\\\nn +\text{(finite terms)}.
\end{eqnarray}

In conclusion, in both cases, pion and nucleon, the rapidity divergence is proportional to the tree-level matrix element. This is expected, because the rapidity divergence of a TMD operator is multiplicatively renormalizable \cite{Vladimirov:2017ksc}. Similarly to QCD, it is independent of the Dirac structure. This is valid at all loop orders, since the rapidity divergence arises in the interaction with the Wilson line, and the Dirac matrix $\Gamma$ does not enter the loop part of the diagram. Similarly, it is independent of the direction of the Wilson lines (i.e., of the sign of $\delta$, in the current form). However, in contrast to QCD, it depends on the flavor (isospin channel) of the operator. We discuss this feature in sec.~\ref{sec:discussion}.

\section{Soft factors in ChPT}
\label{sec:sf}

In QCD, the renormalization factor for rapidity divergences is usually defined via the soft factor, such that
\begin{eqnarray}\label{TMD-renorm}
O_{11}(z,b;\mu,\zeta)=Z^2_{\text{UV}}(\mu,\epsilon)\frac{1}{\sqrt{S(b,\zeta;\epsilon,\delta)}} O_{11}(z,b;\epsilon,\delta),
\end{eqnarray}
where $Z_{\text{UV}}$ is the UV renormalization factor of the operator $U_1$, and $S$ is the soft factor. In this expression, we have indicated the regulators of the UV divergences, $\epsilon$, and of the rapidity divergences, $\delta$. 

The soft factor is conventionally defined as the vacuum matrix element of the eikonalized operator  \cite{Ji:2004wu, Idilbi:2007ff, Echevarria:2011epo, Collins:2011zzd}, i.e., the operator with the parton fields replaced by the corresponding Wilson lines. In the same manner, we define the soft factors in ChPT. The only difference from QCD is that QCD allows only color-singlet operators, while in ChPT isospin indices are allowed. Therefore, the soft factor carries the isospin representation index. In the nucleon case, we define
\begin{eqnarray}
S^I_N(b)&=&\frac{1}{2d_I}\langle 0|\tr\[\tau^A \Pi_R[\infty \bar n+b,b]\Pi_R[b,\infty n+b] \tau^A \Pi_R[\infty n,0]\Pi_R[0,\infty \bar n]\]|0\rangle,
\end{eqnarray}
where $A=0$ (i.e. $\tau^0=\mathbb{I}$) for the $I=0$ case, and $A=1,2,3$ for the $I=1$ case. The factor $d_I=\frac{1}{2}\tr(\tau^A\tau^A)$ is the dimension of the corresponding representation, i.e. $d_{I=0}=1$ and $d_{I=1}=3$. Note that the operator is made of $\Pi_R$, but it is equal to the one made of $\Pi_L$, because the difference between these operators is odd in pion fields, and thus vanishes in the vacuum matrix element.

For the pion-in-pion TMD operator, the situation is more cumbersome, because the pion transforms in the adjoint representation of SU(2). Therefore, its eikonalized counterpart is a Wilson line in the adjoint representation. It is defined as
\begin{eqnarray}
\mathfrak{N}^{ab}[x_1,x_2]=\frac{1}{2}\tr\(\tau^a \Pi_R[x_1,x_2] \tau^b \Pi_R[x_2,x_1] \).
\end{eqnarray}
Then the eikonal limit of the pion matrix elements (\ref{pi:I=0}, \ref{pi:I=1}) is
\begin{eqnarray}
S^{I=0}_\pi(b)&=&\frac{\delta^{ab}}{6}\langle 0|\tr\[\tau^{b'}\tau^{a'}\mathfrak{N}^{a' a}[b,\infty \bar n+b] \Pi_R[b,\infty n+b] \, \Pi_R[\infty n,0] \mathfrak{N}^{\dagger b b'}[0,\infty \bar n]\]|0\rangle,
\\
S^{I=1}_\pi(b)&=&\frac{-i\epsilon^{abc}}{4}\langle 0|\tr\[\tau^{b'}\tau^c \tau^{a'}\, \mathfrak{N}^{a'a}[b,\infty \bar n+b] \Pi_R[b,\infty n+b] \, \Pi_R[\infty n,0] \mathfrak{N}^{\dagger b b'}[0,\infty \bar n]\]|0\rangle.
\end{eqnarray}
In some sense, these are soft factors in a mixed representation, i.e., they couple the fundamental $\Pi$ and the adjoint $\mathfrak{N}$ Wilson lines. Such a curious combination appears because the quark operator carries exclusively fundamental quantum numbers, while the pion carries adjoint ones.

\begin{figure}[t]
\begin{center}
\includegraphics[width=0.46\textwidth]{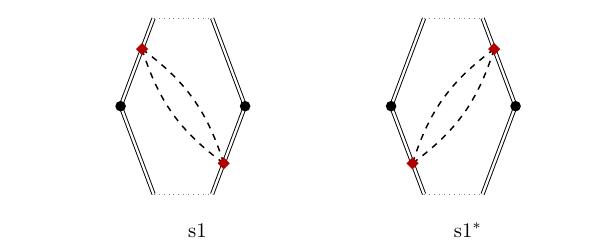}
\end{center}
\caption{\label{fig:soft} The diagrams that produce the leading rapidity divergence in the soft factors.}
\end{figure}

The computation of the soft-factor matrix elements is straightforward. At one loop, the soft factor has no rapidity divergence, since all $b$-dependent diagrams have propagators attached to fixed points. At two loops, there are two rapidity-divergent diagrams, shown in fig.~\ref{fig:soft}. All soft factors are proportional to the same integral and differ only in the group factors. The loop integral reduces to the evaluation of
\begin{eqnarray}
I_S(b)&=&-i\int \frac{d^dl}{(2\pi)^{d}}\frac{e^{i(l\cdot b)} B(l^2)}{(l^+\pm i\delta^+)(l^-+i\delta^-)},
\end{eqnarray}
which is equal to
\begin{eqnarray}
I_S(b)&=&\frac{-2m_\pi^4}{(2\pi)^4} \mathcal{K}(b^2)\ln(\delta^+ \delta^- b^2)+\text{(finite terms)},
\end{eqnarray}
with the functions $B$ and $\mathcal{K}$ defined in (\ref{def:B}, \ref{def:K}). The parameter $\delta^-$ regularizes the rapidity divergence along the $\bar n$ direction. The expressions for the soft factors read
\begin{eqnarray}\label{SF-nuc}
S_N^I(b)&=&1-c_I\frac{4(1-a^2)^2m_\pi^4}{(4\pi F_\pi)^4}\mathcal{K}(b^2) \ln(\delta^+ \delta^- b^2)+\text{(finite terms)}+\mathcal{O}\(\frac{m^6_\pi}{F_\pi^6}\),
\\\label{SF-pion}
S_\pi^I(b)&=&1-\frac{16(1-a^2)^2m_\pi^4}{(4\pi F_\pi)^4}\mathcal{K}(b^2) \ln(\delta^+ \delta^- b^2)+\text{(finite terms)}+\mathcal{O}\(\frac{m^6_\pi}{F_\pi^6}\),
\end{eqnarray} 
where $c_I$ is defined in (\ref{def:cI}).

Combining the TMD operators with the soft factors as in (\ref{TMD-renorm}), we find that the rapidity divergences cancel only for particular values of the parameter $a$. 
\begin{itemize}
\item \textbf{The nucleon case} (compare (\ref{rap-div-nuc}) with (\ref{SF-nuc})): the cancellation requires $|a|=a_N=g_A$. This choice corresponds to the Wilson line representing the eikonalized nucleon propagator.
\item \textbf{The pion case} (compare (\ref{rap-div-pion}) with (\ref{SF-pion})): the cancellation requires $(1-a^2)^2=(1-a^2)$. It is satisfied either by $a=a_\pi=0$ or by $a=a_\pi=\pm1$. The latter corresponds to the trivial case of absent Wilson lines (\ref{W-trivial}) and of rapidity divergences.
\end{itemize}
Assigning $a$ by these rules, one finds that
\begin{eqnarray}\label{TMD-final}
O^I_{h}(z,b;\mu,\zeta)=\[1+\alpha_h^I \frac{2m^4_\pi}{(4\pi F_\pi)^4}\mathcal{K}(b^2)\ln\(\zeta b^2\)+\mathcal{O}\(\frac{m^6_\pi}{F_\pi^6}\)\]O_{11,h}(z,b)+\text{(finite terms)}.
\end{eqnarray}
Here, $\zeta$ is the scale of rapidity evolution, which in the $\delta$-regularization is defined as $\zeta=2(p^+)^2\delta^-/\delta^+$. The constant $\alpha_h^I$ is
\begin{eqnarray}\label{def:alpha}
\alpha_h^I=
\left\{
\begin{array}{lll}
3(1-g_A^2)^2,& I=0, & h=N,
\\
-(1-g_A^2)^2,& I=1, & h=N,
\\
4,& I=0,1,~~ & h=\pi,
\end{array}\right.
\end{eqnarray}
where $a_\pi=0$ is used as the default case. Selecting $a_\pi=\pm 1$, one gets $\alpha_\pi^I=0$.

\section{Discussion and conclusions}
\label{sec:discussion}

In this work, we have demonstrated how to construct the semi-compact building blocks in ChPT, and have constructed analogs of TMD operators at the leading chiral order. The obtained operators have the same geometric structure as the operators in QCD, i.e., parton fields (in the ChPT case, nucleon or pion fields) with an attached staple of Wilson lines. In the twist counting, they have TMD-twist-(1,1), and thus obey the same evolution equations. Each of them is accompanied by a unique generating function of LECs, which could be interpreted as TMD distributions in the chiral limit. We have explicitly constructed the operators for all eight nucleon and two pion TMDPDFs. 

The only exception to the common pattern for TMD operators is the operator corresponding to the Boer-Mulders function in the pion. To fulfill the chiral-odd criterion, it requires the insertion of extra pion fields at the light-cone infinity. Due to this, its tree-level contribution vanishes and it appears chirally suppressed. It could be seen as evidence of an extra smallness of the pion Boer-Mulders TMDPDF. However, such a statement is not conclusive, because the values of the generating functions of LECs remain unknown. In this respect, a direct investigation of TMD distributions in ChPT has little practical importance.

The most important observation and the main result of this work is the demonstration and explicit evaluation of the rapidity divergences of TMD operators in ChPT. They appear in a manner very similar to QCD, i.e., due to the interaction of pions at the ``far end'' of the Wilson line with the bulk of the operator. As in QCD, they can be renormalized by multiplying by the inverse square root of the soft factor, which is defined as the eikonalized version of the TMD matrix element. This leads to the expression (\ref{TMD-final}), which is the main computational result of this work.

Differentiating expression (\ref{TMD-final}) with respect to $\zeta$, one finds the expression for the CS kernel (\ref{evol:zeta}) in ChPT. It reads
\begin{eqnarray}\label{CS-final}
\mathcal{D}_h^I(b,\mu)=-\alpha_h^I \frac{2m^4_\pi}{(4\pi F_\pi)^4}\mathcal{K}(b^2)+\mathcal{O}\(\frac{m^6_\pi}{F_\pi^6}\),
\end{eqnarray}
where $\mathcal{K}$ and $\alpha_h^I$ are defined in (\ref{def:K}) and (\ref{def:alpha}), respectively. This expression can be considered as a model-independent prediction for the CS kernel at a low renormalization scale $\mu\sim m_\pi$ and large $b$, $b\gg (4\pi F_\pi)^{-1}\sim 1$~GeV$^{-1}$.  

The evolution of the CS kernel is given by the equation
\begin{eqnarray}\label{CS:evolution}
\mu \frac{d\mathcal{D}(b,\mu)}{d\mu}=\Gamma(\mu),
\end{eqnarray}
where $\Gamma(\mu)$ is the cusp anomalous dimension in QCD. It is also the cusp anomalous dimension in ChPT, and should be computed from the soft factor. Our computation shows that in ChPT
\begin{eqnarray}
\Gamma(\mu)\sim \mathcal{O}\(\frac{m^6_\pi}{F_\pi^6}\).
\end{eqnarray}
The important point is that $\Gamma$ is independent of $b$ at all orders of ChPT and QCD. Thus, the evolution of the CS kernel (\ref{CS:evolution}) does not change its shape, but only shifts it by a common constant
\begin{eqnarray}
\mathcal{D}(b,\mu)=\mathcal{D}(b,\mu_0)+\int_{\mu_0}^\mu \frac{d\mu}{\mu}\Gamma(\mu).
\end{eqnarray}
Therefore, the large-$b$ shape of the CS kernels predicted in (\ref{CS-final}) remains valid also at high-energy scales.

In this way, the CS kernel is fixed by QCD computations at small values of $b$ \cite{Vladimirov:2016dll, Duhr:2022yyp, Moult:2022xzt}, $b\ll \Lambda^{-1}_\text{QCD}\sim 1$~GeV$^{-1}$, and (up to an additive constant) by ChPT (\ref{CS-final}) at large values of $b$, $b\gg (4\pi F_\pi)^{-1}\sim 1$~GeV$^{-1}$. In between these regimes, the strong interaction undergoes the confinement transition, and thus perturbative computations cannot be applied. In the intermediate-$b$ regime, the characteristics of the CS kernel change, in comparison to the usual QCD assumptions. In particular, we observe that 
\begin{itemize}
\item At large $b$, the CS kernel is independent of the polarization structure of the TMD operator, as is the QCD prediction for small $b$. Therefore, we can conclude that the CS kernel is the same for all eight TMDPDFs.
\item At large $b$, the CS kernel depends on the parton's flavor and on the type of hadron. This contradicts the perturbative QCD observation that the CS kernel is flavor/hadron independent. However, this observation is consistent with the common picture, in which the CS kernel describes the strong interactions between partons of the colliding high-energy hadrons. In this picture, at small distances, where the hadrons overlap, such interaction is gluon dominated, and thus flavor and hadron blind. At large distances, where the hadrons are well separated, the strong interactions hadronize into pions, and thus become isospin and hadron dependent.
\item At large $b$, the CS kernel depends on the parton's flavors via the isospin couplings. It implies that the evolution of pure-flavor channels is mixed, in contrast to QCD. The eigenstate combinations $u\pm d$ remain autonomous.
\item The anomalous dimension $\Gamma$ of the CS kernel (\ref{CS:evolution}) at very low energy is flavor independent. However, it is hadron dependent because the chiral Wilson lines are hadron dependent via the parameter $a$.
\item In both QCD and ChPT, the CS kernel is process independent, i.e., the SIDIS and Drell-Yan definitions have the same rapidity divergences.
\end{itemize}

However, the deviations from the QCD expectations should not be considered very strong, because
\begin{eqnarray}\label{CS-large-b}
\mathcal{D}_h^I(b,\mu)\sim-\alpha_h^I \frac{\pi}{2}\frac{m_\pi}{b^3(4\pi F_\pi)^4}e^{-2m_\pi b}
\end{eqnarray}
at large values of $b$, in agreement with the expectations of ref.~\cite{Collins:2014jpa}. 
It means that at large $b$, the CS kernel rapidly approaches a constant. The dependence on the flavor is very small and invisible at current precision, although not absolutely negligible. In particular, the difference between the $I=1$ and $I=0$ cases for the proton is $0.03$ at $b=2$~GeV$^{-1}$ and $0.006$ at $b=3$~GeV$^{-1}$.

It is also instructive to inspect the chiral limit of (\ref{CS-final}). In this limit, it reads
\begin{eqnarray}
\mathcal{D}_h^I(b,\mu)\sim-\alpha_h^I \frac{2}{(4\pi F_\pi)^4}\(\frac{1}{3 b^4}-\frac{m_\pi^2}{2b^2}+...\).
\end{eqnarray}
The leading term of the chiral expansion is pion-mass independent. Therefore, in the regime $(4\pi F_\pi)^{-1} \ll b \ll m_\pi^{-1}$, the value of the CS kernel depends weakly on the mass of the pion. Moreover, since at larger $b$ the CS kernel approaches a constant (\ref{CS-large-b}), the very weak $m_\pi$ dependence can be safely extended to $b\gg (4\pi F_\pi)^{-1}$ (or $b\gtrsim 0.5$fm). This makes this regime safe for lattice simulations, which depend on the mass of the pion. This observation agrees with ref.~\cite{Tan:2025ofx}, which investigated the dependence on the pion mass and found it to be negligible.

\begin{figure}[t]
\begin{center}
\includegraphics[width=0.46\textwidth]{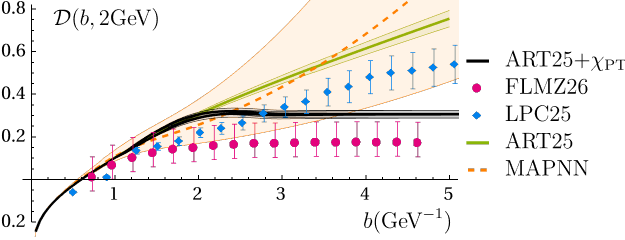}
\end{center}
\caption{\label{fig:CS} Example of a possible behavior of the CS kernel (black) in comparison to recent determinations of the CS kernel. This curve is an interpolation (in the range $(1,3)$~GeV$^{-1}$) between the ART25 curve and the ChPT computation (\ref{CS-final}). Lines show the extractions of the CS kernel from data by ART25 \cite{Moos:2025sal} and MAPNN \cite{Bacchetta:2025ara}. The points show recent lattice extractions by FLMZ26 \cite{Francis:2026czb} and LPC25 \cite{Tan:2025ofx}.}
\end{figure}

In fig.~\ref{fig:CS}, we demonstrate a possible behavior of the CS kernel as an interpolation between the ART25 extraction \cite{Moos:2025sal} and the expression (\ref{CS-final}). The black curve in fig.~\ref{fig:CS} contains both $I=0$ and $I=1$ cases. These cases, however, are indistinguishable at this scale. We emphasize that this curve is a result of a simple interpolation, and the real curve can be different, although according to (\ref{CS-final}) it should eventually turn into a constant.

\acknowledgements

Some ideas used in this work stem from discussions with the late Maxim Polyakov, which took place in 2019, and to whom the author remains very grateful. The author is funded by the Atracci\'on de Talento Investigador program of the Comunidad de Madrid (Spain) No. 2020-T1/TIC-20204. During the preparation of this manuscript, the author utilized Claude Opus 5.5 developed by Anthropic to verify several algebraic relations and to review the text for grammatical clarity. After using this tool, the author reviewed and edited the content and takes full responsibility for the final manuscript. 

\bibliography{bibFILE}
\end{document}